\documentclass[aps,epsfig,showpacs,floats,twocolumn]{revtex4}
\usepackage{epsfig}
\usepackage{amssymb}
\newcommand{\moge}{$a$-Mo$_{2.7}$Ge}

\begin{document}

\title{Vortex relaxation and coupling in superconducting
heterostructures studied by STM}
\author{G.J.C. van Baarle, F. Galli, P. H. Kes and J. Aarts}
\affiliation{Kamerlingh Onnes Laboratory, Leiden University, P.O. Box 9504, 2300 RA
Leiden, The Netherlands}
\date{\today}

\begin{abstract}
In a sandwich consisting of two superconducting films, one weakly pinning and one
strongly pinning, the vortex positions in both films are determined by the strongly
pinning material and the vortex lattice is disordered in both films. We used
(strongly pinning) NbN and (weakly pinning) \moge~and studied, by directly imaging
the vortex core positions with a scanning tunnelling microscope, how this disorder
is restored with increasing thickness of \moge~layer or when the interface is made
insulating. For clean interfaces we find that the first reordering of the vortex
lattice is found at a layer thickness wich is compatible with the first bending mode
of the vortex lines. Making the interface insulating we find that order is restored
quickly. We argue that this is can be understood from the competition between the
Josephson force working on the vortex segments on the one hand, and the elastic
restoring forces inside the weakly pinning layer on the other hand.
\end{abstract}

\maketitle

\section{Introduction}
In the investigation of the properties of vortices and vortex lattices in type-II
superconductors, direct imaging of the vortex positions at the surface of the
material can be a useful tool. For magnetic fields of practical interest the
distance between vortex lines becomes small (roughly 50~nm at 1~T) and Scanning
Tunnelling Microscopy (STM) gains an advantage over magnetic imaging techniques
since STM is only sensitive to the vortex cores. A disadvantage of the technique is
that the sample surface needs to be clean and conducting. For crystals, this can be
done by breaking or cleaving, which has been successful for such materials as
NbSe$_2$ \cite{hess89,troyan99} or Bi$_2$Ca$_2$SrCu$_2$O$_{8-\delta}$
\cite{renner98,fischer06}. For thin films no such simple preparation method exists
and since surfaces easily oxidize, reports on vortex imaging on superconducting
films are relatively few. One way to overcome this problem is to protect the film
surface with a thin capping layer of a noble metal such as Au. As we showed before,
this works well on films of \moge, a weakly pinning superconductor, capped with 3~nm
Au and clear images of the vortex structure could be obtained by mapping the
proximity-induced gap at the Au surface \cite{baarle03}. The Au layer should be
flat, which means that Au should be wetting the surface during deposition, and the
interface should be transparent. Neither condition is always met. For instance,
although a reasonably flat Au film forms on films of strongly pinning NbN,
spectroscopy shows very unclear vortex images, presumably because the rough NbN
surface causes a bad NbN / Au interface. This can be overcome by depositing a thin
(20 nm) film of \moge~followed by the Au protection layer. Clear vortex contrast
could again be obtained, and showed the NbN vortex lattice (VL) to be completely
disordered \cite{baarle03}. \\
This use of a \moge/Au template layer assumes that the vortex  positions as fixed by
the NbN layer are transferred in true fashion to the Au surface, which should be the
case when the \moge~layer is too thin to yield appreciable bending or rearrangement
of the vortex cores. In this work we further address this issue, which consists of
two parts. The first involves the vortex-vortex interactions within the template
layer, which might lead to reorientation of the vortices in that layer or, more
specifically, to the recurrence of the well-ordered \moge~VL. The second question is
with respect to the coupling of the vortices in the two superconducting layers. If
the interface has a metallic character, the proximity effect will play a major role
in coupling the vortices; if the surface of the host superconductor is insulating,
as is often the case in oxide materials, the coupling may become of the Josephson
type, or merely electromagnetic, and this can allow freedom for the vortex in the
template to move with respect to the one in the host. \\
The structure of the paper is as follows. We first introduce the experimental
techniques and the data analysis process, in particular the use of the
autocorrelation function. Then we address the question of the length scale on which
the disorder injected by the NbN layer is healed in the \moge~template layer, by
measuring samples with different (template) thicknesses $d_t$ of the \moge-layer. We
show data on the variation of the autocorrelation functions as function of $d_t$,
and we discuss them by considering the bending modes of vortex lines in \moge. Then
we present experimental results on the issue of coupling, by studying
heterostructure samples with a badly conducting or insulating interface layer. In
the discussion, we consider the strength of the electromagnetic and Josephson
coupling forces which couple the vortex segments in the two superconducting layers.
This coupling is in competition with the inter-vortex forces within the template
layer, which tend to reorient the vortices to their triangular arrangement.

\section{Experimental details and data processing}
In this work \moge~is used as a template layer because of its suitable properties.
With a Ginzburg-Landau parameter $\kappa \approx 100$ it is a strong type-II
superconductor, while its amorphous structure makes it a relatively weakly pinning
material. Furthermore, its large London penetration depth $\lambda_L$ yields a small
tilt modulus $c_{44}$, which means that the VL retains a two-dimensional (2D)
character up to quite large thicknesses, as will be discussed below. Finally,
amorphous materials show very flat surface characteristics. The \moge~layers were
rf-sputtered from a composite Mo/Ge target in Ar atmosphere. The film thickness
ranged from 25~nm to 100~nm. When necessary, we denote the thickness by \moge(x),
with x the thickness in nm. During sputtering the substrate was water-cooled and
after each minute of sputtering, and interruption of 1 minute was made to prevent
heating and possible recrystallization of the films. The sputtering rate was chosen
as $\sim ~8$~nm/minute. The superconducting transition temperature $T_c$ was 5.7~K
for the 25~nm films, increasing up to $~6.3$~K for $d_t = 50$~nm, after which $T_c$
became constant. The critical current for onset of vortex motion for a 500~nm thick
film was found to be or the order of $10^6$~A/m$^2$. Also relevant is the value of
the critical field $B_{c2}$ at the measurement temperature of 4.2~K, which was about
4.2~T for $d_t$ = 25~nm, increasing to 5.2~T for $d_t$ = 50~nm and beyond. The STM
measurements were performed in fields up to 1~T, which means maximum reduced fields
$b = B/B_{c2} \approx$ 0.2 - 0.3.
\\
The strongly pinning NbN films ($J_c > 10^8$ A/m$^2$) were reactively rf-sputtered
from a Nb target in an Ar/N$_2$ atmosphere on Si substrates. Throughout the study,
we used a thickness of 67~nm. The films are polycrystalline, as is known from high
resolution transmission electron microscopy studies \cite{pruymboom87}. The
rms-roughness of the films was determined by AFM to be 1.5~nm, measured on a fresh
NbN surface area of 1~$\times$~1~$\mu$m. The films showed a $T_c$ of 11.4~K and a
superconducting coherence length $\xi=4.7$~nm. \\
Protective Au films were rf-sputtered in two steps. First, about 2~nm was sputtered
in a pure Ar atmosphere; next, oxygen (partial pressure 5~\%) was mixed in the
sputtering gas, and another layer of 2~nm was sputtered. This O$_2$-addition
promotes the formation of smaller Au grains which improves the resolution of the
measurements. \\
%
% proefschrift fig.4.9 b/c
%
\begin{figure}
\begin{center}
\includegraphics[width=8.5cm]{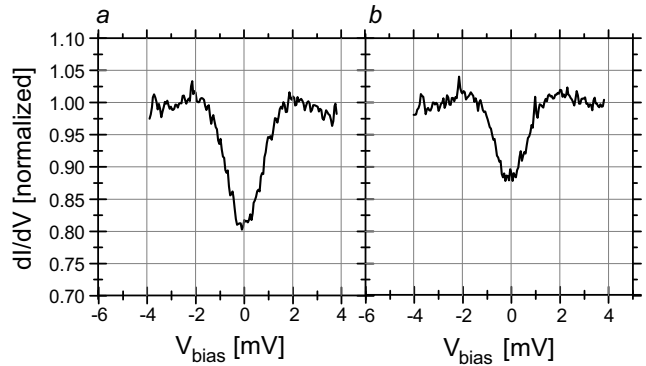}
\caption{Averaged normalized derivatives $d$I/$d$V for two samples, acquired at T =
4.2~K and B = 0.3~T. (a) Sample NbN/MoGe(50)/Au(6); (b) Sample
NbN/MoGe(50)/Au(10)}\label{fig1:gaps}
\end{center}
\end{figure}
The STM-setup was the same as in a previous report \cite{troyan99}, with the sample
immersed in liquid He after it was mounted in air. In this procedure field-cooled
measurements are not possible, since the level of the liquid is higher than the
magnet. All data were therefore taken by zero-field-cooling the sample, and then
increasing the magnetic field. For all experiments we employed full
current(I)-voltage(V) spectroscopy, with a single curve consisting of 200-300
points. The voltage range was chosen such that a good fit could be made to the Ohmic
resistance outside the gap region, in practise typically $\pm 4$~mV. Several curves
were taken per pixel, and images typically contained 128 $\times$ 128 pixels. The
total acquisition time for such an image was of the order of 30 minutes to 1 hour,
depending on the surface quality. The maximum scan size, for the given grid, is
determined by the vortex core size. The vortex signature is visible in a region with
a diameter of roughly 4~$\xi$, and needs about 3~pixels in this region for a proper
determination. This yields 7~nm per pixel, and a scan size of maximally 0.9~$\mu$m.
The vortex positions were found by taking the ratio of the zero-bias and high-bias
slopes, as obtained by numerical differentiation. Theoretically, this is a value
between 0 (full gap) and 1 (no gap). In practise the gap is not well developed due
to the finite temperature and the proximizing Au layer and in the gap regime the
ratio is of the order 0.8. The resulting range (0.8 to 1) is then translated into a
grey scale. Fig.~\ref{fig1:gaps} shows the effect of the thickness of the Au layer
by comparing the gaps for two samples NbN/MoGe(50)/Au, where the first one has a Au
layer thickness of 6~nm, while the Au layer of the second one is almost two times thicker. \\
%
% proefschrift fig. 3.7a-d
%
\begin{figure}
\includegraphics[width=8cm]{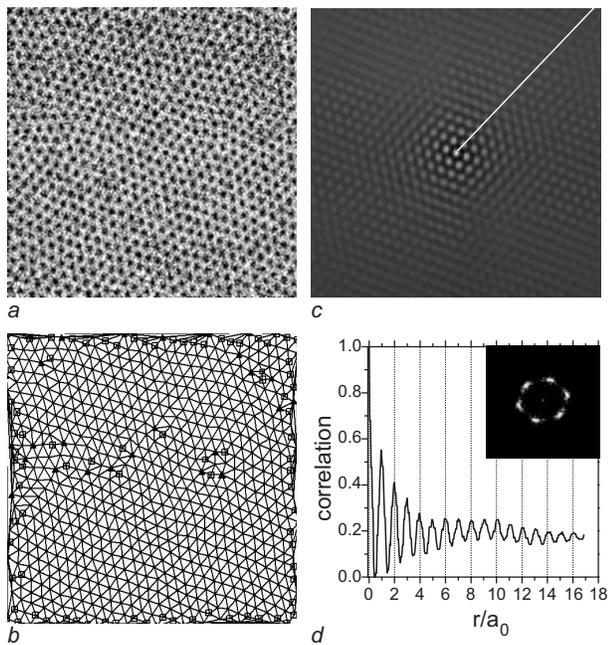}
\caption{(a) Vortex lattice (VL) image on a sample MoGe(48)/Au, acquired at T =
4.2~K and B = 0.8~T. The scan size is 1.2~$\mu$m $\times$ 1.2~$\mu$m. The raw data
were smoothed with a Gaussian filter of 1 pixel width; (b) Triangulation
representation of the VL. The lines represent bonds, the symbols denote deviations
from the 6-fold coordination of the perfect trangular lattice (triangles (squares) :
7-fold (5-fold) coordination); (c) Normalized autocorrelation map. (d) Cross-section
of the autocorrelation map along the white line in (c). The inset gives the central
part of the 2D-Fourier transform of the direct-space image given in (a)}
\label{fig2:VL-autocorr}
\end{figure}
A useful way of determining the amount of correlation of vortex positions is to
compute the two-dimensional autocorrelation function $G(\bf{r})$ for a given image,
defined as $G(\bf{r}) = <\rho(\bf{r}_1)\rho(\bf{r}_1+\bf{r})>$. The function
computes the overlap of the image with a copy of itself which is displaced by a
vector $\bf{r}$. In the actual calculation, the result is normalised in order to
correct for the decreasing amount of datapoints available with increasing $r$.
Fig.~\ref{fig2:VL-autocorr} gives the result of this procedure for a VL imaged on a
\moge-film of 48~nm thick at an applied field $B$ of 0.8~T. Triangulation
(Fig.~\ref{fig2:VL-autocorr}b) shows that the VL is quite free of defects. The
autocorrelation map of $G(\bf{r})$ (Fig.~\ref{fig2:VL-autocorr}c) shows very little
blurring towards the edges, and makes evident that the VL has both high
translational and rotational bond order. By making a cross-section along a direction
from the center of the image through one of the points of the hexagon, the loss of
correlations can be quantified. (Fig.~\ref{fig2:VL-autocorr}d) shows such a
cross-section, plotted in units $r/a_0$ and normalized to the value at $r/a_0 = 0$.
Here, $a_0$ the intervortex distance given by $a_0 = 1.07 \sqrt{\Phi_0} / B$. It
appears that $r_c$ (somewhat arbitrarily defined as the displacement vector where
the autocorrelation function does not show lattice periodicity anymore) is larger
than 16~$a_0$ \cite{note-Rc}.

\begin{figure}
\includegraphics[width=7cm]{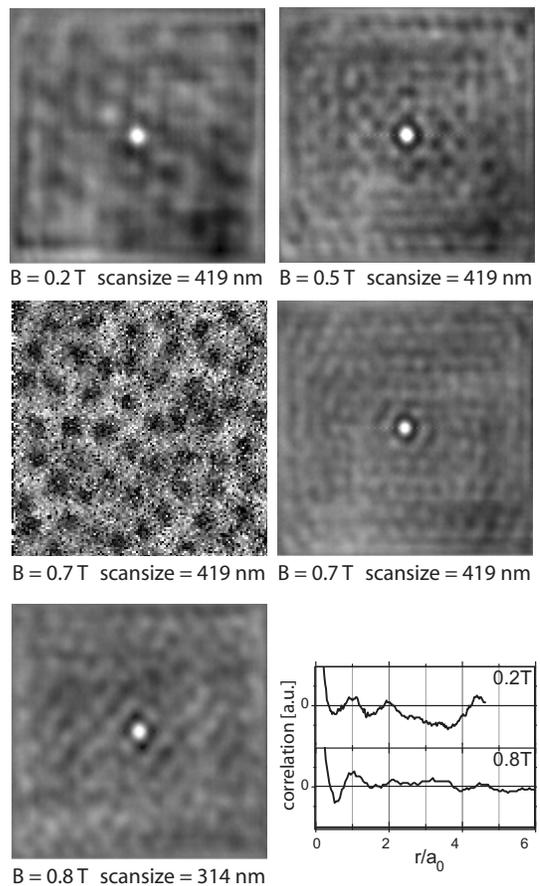}
\caption{Compilation of results of vortex lattice (VL) imaging of a sample
NbN(50)/Au. The middle panel in the left-hand corner shows a VL image acquired at T
= 4.2~K and B = 0.7~T. The adjacent panel shows the normalized autocorrelation map
computed from theses data. Other panels show autocorrelation map at fields of 0.2~T,
0.5~T and 0.8~T. The lower right-hand panel shows cross-sections of the
autocorrelation images taken at 0.2~T and 0.8~T.}\label{fig3:NbN}
\end{figure}

In contrast, the VL in NbN is almost completely disordered, as shown in
Fig.~\ref{fig3:NbN} for a 50~nm film with a Au capping layer. In this case, we find
$r_c \approx 1.5 a_0$ and almost no orientational correlation. Note also the
relatively poor quality of the vortex image, due to the fact that no \moge-template
layer is present.
%
% proefschrift fig. 4.15 (top) plus 2 panels from fig. 4.17
%
Next we shall investigate how this disorder propagates through an
intervening \moge~layer. \\

\section{Healing of the vortex lattice}
\subsection{Experimental data}
In order to investigate on what length scale the VL disorder starts to approach the
\moge~hexagonal lattice again, we recorded vortex images at three field values
(0.2~T or 0.3~T, 0.7~T or 0.8~T, and 1~T) for NbN/\moge/Au samples with $d_t$ =~0,
25, 50 and 100~nm.
%
% proefschrift 4 panels from fig. 4.15 / 16
%
\begin{figure}
\includegraphics[width=8cm]{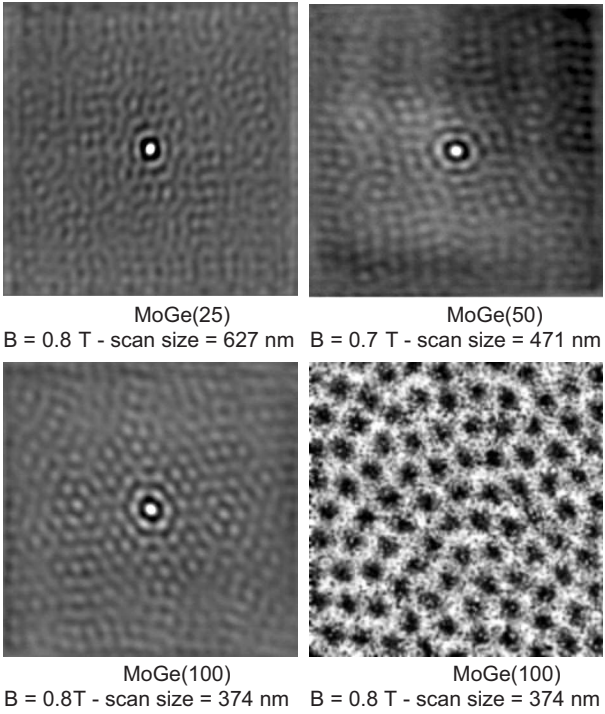}
\caption{Compilation of autocorrelation functions resulting from vortex lattice (VL)
imaging of samples NbN(67)/\moge(d$_t$)/Au, with $d_t$ = 25~nm, 50~nm, 100~nm
acquired at T = 4.2~K and fields of either B = 0.7~T or 0.8~T as indicated. The
lower right-hand corner shows real space data.}\label{fig4:autocorr-d}
\end{figure}
In Fig.~\ref{fig4:autocorr-d} we show the autocorrelation (acr) results computed
from the raw vortex images as measured in either $B=0.7$ or $B=0.8$~T, for the
samples with $d_t$ = 0, 25, 100~nm, together with one real image for $d_t$ = 50~nm
as a sample of the data quality. Since the contrast and the noise-level of the data
are affected by the varying experimental conditions the relevance lies in the the
periodicity of the structures. The gray-scale of the acr-images is chosen such that
the structure of the data is clearly visible. From the data it appears that for the
thinnest \moge~layer, the vortex positions are basically unchanged. The first ring
at $r=a_0$ stands out more clearly, but that is simply due to the better data
quality. There is also some increased intensity at the hexagon corners, indicating
the onset of orientational order. On increasing the thickness of the \moge~layer we
find that both the transversal and orientational order develops more and more. This
is better followed in the cross-sections of the acr-images presented in
Fig.~\ref{fig5:cross-d}.
%
% proefschrift fig. 4.17
%
\begin{figure}[t]
\includegraphics[width=8.5cm]{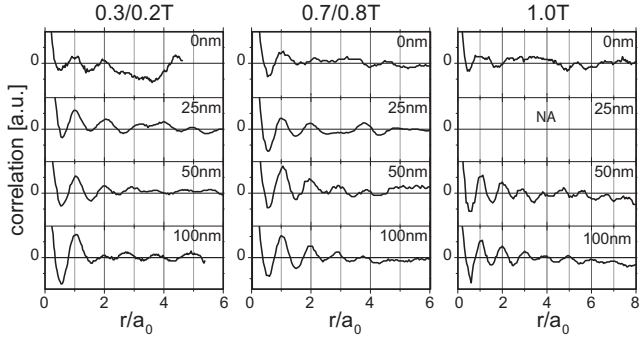}
\caption{Cross-sections from the autocorrelation maps of vortex lattices measured on
samples NbN(67)/\moge(d$_t$)/Au, with $d_t$ = 0~nm, 25~nm, 50~nm, and 100~nm
acquired at T = 4.2~K and fields of either B = 0.7~T or 0.8~T as indicated.
}\label{fig5:cross-d}
\end{figure}
Clearly, the reordering occurs but slowly. For $d_t$ = 25~nm, $r_c / a_0 \approx 2$
at all fields. For 100~nm, this has increased to about 4 (in the 0.8~T data) or
maybe 6 (in the 1~T data). Although the signature for healing therefore is
unequivocally present, the order at a \moge~thickness of 100~nm is still far less
than in the unconstrained vortex lattice.

\subsection{Bending modes in \moge.}
\begin{figure}
\includegraphics[width=6cm]{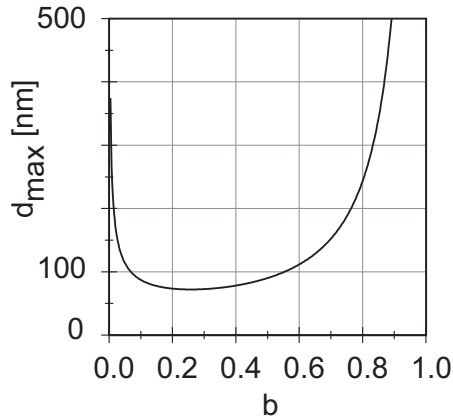}
\caption{Dependence of $d_{max}$, the maximum film thickness for which (tilt)
deformations are absent in \moge-films, on the reduced magnetic field $b$}
\label{fig6:bending}
\end{figure}

In order to understand this length scale, we next consider the length over which a
vortex line can bend. This involves both tilting the line and shearing it with
respect to the whole VL. Starting from a 2D collectively pinned VL, we can estimate
the film thickness where the first bending starts to be relevant. The extent to
which deformations can develop in the transverse and longitudinal direction depends
on the balance of the tilt and shear energy densities $E_{s,t}$ which are given in
the elastic limit by $E_s=\frac{1}{2} c_{66}(u/s)^2$ for shear deformations (with
$c_{66}$ the shear modulus of the VL) and $E_t=\frac{1}{2} c_{44}(u/l)^2$ (with
$c_{44}$ the tilt modulus of the VL) for tilt deformations of a vortex in a medium
where $s$ and $l$ respectively are the transversal and longitudinal axis of the
cigar-shaped surface with constant displacement $u$ (with $u\ll a_0$ in order to be
in the linear regime). The energy balance leads to a relation for the ratio between
$s$ and $l$:
\begin{equation} \label{elasticbalance}
l = s \sqrt{\frac{c_{44}(k_{\perp},k_{\parallel})}{c_{66}}}
\end{equation}
This means that if we set the typical length scale for longitudinal deformations to
the film thickness ($l = d$), the typical transverse length scale $s$ is set by the
square root of the ratio of the elastic constants. The smallest transverse length
scale is simply given by $s=a_0$. The thickness $d_{max}$ where these deformations
start to be relevant can be obtained by neglecting the very small dispersion of
$c_{66}$ and taking for $c_{44}=c_{44}(k_{\perp},k_{\parallel})$:
\begin{equation}\label{c44}
c_{44}(k_{\perp},k_{\parallel})=
\frac{c_{44}(0)}{1+\lambda^2(k_{\perp}^2+k_{\parallel}^2)} \approx
\frac{c_{44}(0)}{\lambda^2(k_{\perp}^2+k_{\parallel}^2)},
\end{equation}
where $c_{44}(0)=B^2/\mu_{0}$; in the last step we made use of the fact that
$\lambda \gg a_0$, as well as that $b \le 0.3$ (making a field correction for
$\lambda$ unnecessary). Since $\lambda$ is very large, we use the non-local
expression. The wavevector $k_{\perp}$ for tilt deformations runs from 0 to $K_0$,
the Brillouin zone boundary. In a circular approximation the Brillouin zone boundary
is given by $K_0 = (4\pi^2 B/\Phi_0)^{1/2}$. In principle one should integrate over
all the values for $k_{\perp}$, but from eq.~\ref{c44} we find that the large
$k$-contributions dominate, leading to the approximation:
\begin{equation}
c_{44}(k_{\perp},k_{\parallel})\approx \frac{c_{44}(0)}{\lambda^2
\left((\frac{2\pi}{a_0})^2 + k_{\parallel}^2 \right)}.
\end{equation}
In case of a given vortex length $d$ the wavevector $k_{\parallel}$ for tilt
deformations is quantized: $k_{\parallel}=n\pi/d$ with $n\in\mathbb{N}$. Since we
are looking for the case that the vortex line just starts to bend, we are interested
in the case where the $n=1$ mode becomes relevant, which leads to the expression
\begin{equation}
c_{44}(k_{\perp},k_{\parallel})\approx \frac{c_{44}(0)}{\lambda^2 \left(
(\frac{2\pi}{a_0})^2 + (\frac{\pi}{d})^2\right )}.
\end{equation}
It can easily be seen that the relevance of higher order terms goes down with
$n^{-2}$. Together with Eq.~\ref{elasticbalance} we have an expression for the
maximum film thickness $d_{max}$ for which deformations of the vortex line by
bending are absent. The dependence of $d_{max}$ on $b$ is plotted in
Fig.~\ref{fig6:bending} using the parameters for \moge. We find that in the
experimental regime of $0.01<b<0.3$ for having a 2D system the thickness of the
\moge ~film should be smaller than
$\sim 75$~nm. \\
Going back to the data, it appears that at 100~nm the first bending mode becomes
relevant, especially for $b > 0.15$. This was not to be expected immediately. For a
vortex lattice with strong disorder the length at which the first bending mode
appears might actually be much larger, due to the strong renormalization of the
elastic constants resulting from the forced disorder. Still, the conclusion is that
apparently this consideration is less important, and the scale on which healing
starts to appear can be estimated by considering the first bending mode, while, on
the other hand, more modes are needed to fully restore the VL.
%
% proefschrift fig.4.8b
%

\section{Interface coupling}
\subsection{Experimental data}
\begin{figure}
\includegraphics[width=8.5cm]{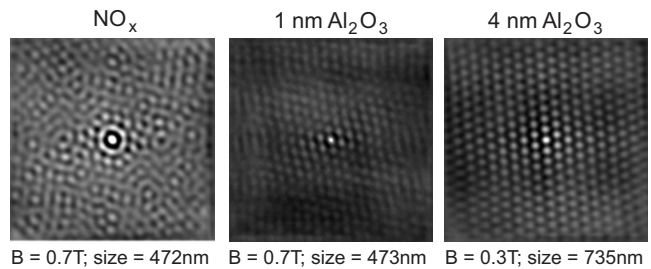}
\caption{Autocorrelation maps from vortex lattices measured at 4.2~K on samples with
different interfaces. (left) NbN(67)/NbNO/\moge(50)/Au, taken at 0.7~T; (middle)
NbN(67)/Al$_2$O$_3$(1)/\moge(50)/Au, taken at 0.7~T; (right)
NbN(67)/Al$_2$O$_3$(4)/\moge(50)/Au, taken at 0.3~T.}\label{fig7:acr-interfaces}
\end{figure}

\begin{figure}
\includegraphics[width=8.5cm]{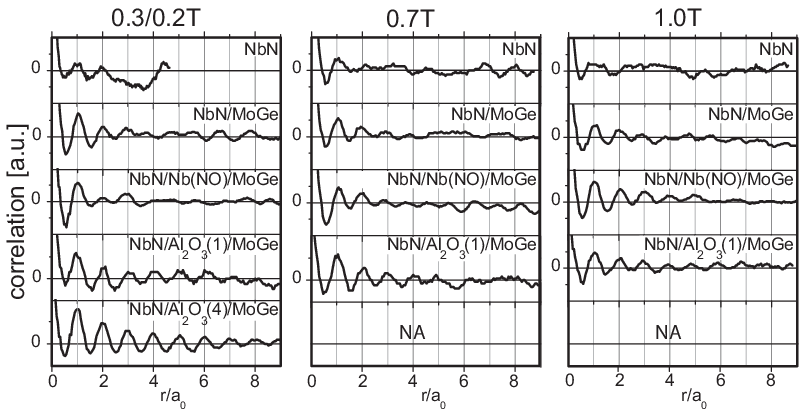}
\caption{Comparison of cross-sections from autocorrelation maps of vortex lattices
measured at 4.2~K on samples NbN(67)/Au, NbN(67)/\moge(50/Au,
NbN(67)/NbNO/\moge(50)/Au, NbN(67)/Al$_2$O$_3$(1)/\moge(50)/Au, and
NbN(67)/Al$_2$O$_3$(4)/\moge(50)/Au, taken at 0.3~T, at different magnetic fields as
indicated. In the left-hand panel, the data on NbN(67)/Au were taken at 0.2~T, all
others at 0.3~T.}\label{fig8:cross-interfaces}
\end{figure}

In order to investigate how the interface between NbN and \moge~affects the vortex
postions, we prepared a variety of samples with a NbN layer of 67~nm, a 50~nm thick
\moge~layer and a $\sim$3.5~nm Au layer, where we gradually reduced the interface
transparency. Next to a reference sample, we first prepared a sample for which the
NbN layer was exposed to pure O$_2$ gas ($\sim 18$~mbar for 15 minutes) prior to the
deposition of the \moge/Au layers. The oxidized layer produced in this way is very
thin; oxidation in air at temperatures of 180~$^{\circ}$C or less is known to
oxidize only the first one or two atomic layers \cite{frankenthal83}. To reduce the
interface transparency even more we deposited an insulating Al$_2$O$_3$ interface
layer of 1~nm or 4~nm thick in between the NbN layer and the \moge~layer.
\\
 In Figs.~\ref{fig7:acr-interfaces},~\ref{fig8:cross-interfaces}
we present a selection of the acr-images and cross-sections as produced from the
raw-data vortex images measure at 4.2~K for various values of $B$.
%
% proefschrift  2x 0.7 T, 1x 0.3 T uit fig.4.11-4.13
%
From the acr-data it can be clearly seen that on changing the interface from
'perfect and thin' to 'insulating and thick' the decoupling of the vortex lattices
in the two superconducting layers rapidly sets in, showing already a clear structure
for the sample with the oxidized NbN surface.
%
% proefschrift fig.4.14
%
From the cross-sections, for the oxided surface we find a value for $r_c/a_0$ of
about 4, for the 1~nm Al$_2$O$_3$ about 7, and for the 4~nm Al$_2$O$_3$ larger than
8 (the maximum for this image size), In other words, the 4~nm insulating interface
has completely decoupled the \moge~vortex lattice from the one in the NbN.
\\
\subsection{Shear force versus Josephson coupling}
To gain some qualitative understanding for this behavior, we assume that the vortex
positions in the template layer are the result of a competition between two forces.
One is the Josephson coupling force, which keeps the vortices in the template in
registry with the those on the other side of the interface. The other is the elastic
force which wants to restore the ordered Abrikosov lattice, and which we equate to
the shear force necessary to displace a vortex line from its equilibrium position in
the lattice. In principle, also the electromagnetic force works to maintain
registry, but a simple estimate shows that the electromagnetic coupling is much
smaller than the Josephson coupling for the very thin interfaces under
consideration. We refer this point to the end of the discussion. \\
Starting with the shear force, for small displacements $u<<a_0$ from the ideal
lattice positions (the elastic limit) the typical force $F_{sh}$ necessary to move a
single vortex line inside the template layer of length $d_t$ over a distance $u$ is
given by
\begin{equation}\label{vortex-vortex}
F_{sh}=c_{66}ud_t = \frac{\Phi_0 B_{c2}(t)ud_t}{16\pi\mu_0\lambda^2}\;
b(1-0.29b)(1-b)^2 ,
\end{equation}
where the extrapolation formula of Brandt \cite{brandt76} is used for the elastic
modulus $c_{66}$. This approach does not take into account that the VL may be
disordered (not in equilibrium), and may contain defects which soften the lattice,
so it should be considered as an upper limit. By rewriting $c_{66} = \Phi_0^2 f(b) /
(16 \pi \mu_0 \lambda^2 a_0^2) $, with $f(b) = (1-0.29b)(1-b)^2$ and using the fact
that $b = B / B_{c2} \approx \Phi_0 / (a_0^2 B_{c2})$, we can also write $F_{sh}$ in
terms of the quantity $\epsilon_0 = \Phi^2_0 / (4 \pi \mu_0 \lambda^2)$ as
\begin{equation} \label{eq:shear}
F_{sh} = \frac{\epsilon_0 u d_t f(b)}{4 a_0^2}.
\end{equation}

For the Josephson coupling force, we start from the expressions for the Josephson
energy $E_J$ as function of a displacement of the vortex segment $u$, given in
ref.~\cite{goldschmidt05} :
\begin{equation}
\frac{\pi E_{J}}{\epsilon_0 d_i} = \left \{ \begin{array}{ll} 0.707 \left(
\frac{\textstyle{u}}{\textstyle{r_J}} \right)^2 \; \ln (
\frac{\textstyle{9r_J}}{\textstyle{u}} ) \;, & u \le 2 r_J
\\ \\
2.828 \left(\frac{\textstyle{u}}{\textstyle{r_J}} \right) - 1.414 \;, & 2 r_J < u
\end{array} \right.
\end{equation}
\\
Here, the Josephson length $r_J$ is given by $r_J = \gamma d_i$, with $d_i$ the
thickness of the insulating layer and $\gamma$ the anisotropy parameter of the
system defined by $\gamma = \lambda_{\perp}/\lambda_{\parallel}$ =
$(m_\perp/m_\parallel)^{1/2}$ = $\xi_\parallel/\xi_\perp$. These expressions for
$E_J$ are derived from fitting the numerical solutions of a model which consist of
two stacks of pancake vortices in an anistropic superconductor (characterised by its
anisotropy parameter $\gamma$), displaced with respect to each other by $u$. For
very small displacements, the expression for $E_J$ is equivalent to an earlier one
given by Koshelev and Vinokur~: $E_J(u) =  \frac{\epsilon_0 d_i}{2}
(\frac{u}{r_J})^2 \ln(r_J/u)$ \cite{koshelev98}. For larger displacements, $E_J$
becomes linear in $u$, which means the Josephson force $F_J$ becomes constant and is
simply given by
\begin{equation}
F_J = \frac{2.8 \; \epsilon_0 }{\pi \gamma} .
\end{equation}
We use this expression as an estimate for the maximum Josephson force
\cite{note-model}. Comparing it to the shear force of Eq.~\ref{eq:shear}, the
condition for $F_{sh} \ge F_J$ becomes
\begin{equation}
\frac{u}{a_0}  \ge \frac{3.6}{\gamma} \frac{a_0}{d_t} \frac{1}{f(b)} .
\end{equation}
Not surprisingly, the result says that the necessary displacement for the shear
force to win from the Josephson coupling force is inversely proportional to the
length of the vortex in the template layer ($d_t$), and also inversely proportional
to the anisotropy parameter. For our experiments, typical values are $d_t$ = 50~nm
and B = 0.3~T, which corresponds to ~$a_0$ = 89~nm and $f(b) \approx$~1, leading to
$(u / a_0 ) \ge 6.4 / \gamma$. Assuming a value of $u / a_0 \approx$ 0.3 for full
disorder, the template layer will become ordered due to shear stress for $\gamma$
$>$~20. This is a very reasonable number for the 4~nm Al$_2$O$_3$ interface layer.
In a study on multilayers of \moge/Ge \cite{white94}, the following values for
$(d_i,\gamma)$ are quoted : (4.5~nm, 7); (5.5~nm, 13), (6.5~nm, 22). The strong
increase with $d_i$ obviously reflects the exponential dependence of the coupling
through an insulating interface, and the numbers given above make it quite
reasonable to expect that decoupling has not yet been achieved for Al$_2$O$_3$(1),
but occurs for Al$_2$O$_3$(4). Also, the result indicates that the field dependence
of the shear modulus is hardly important in comparison with the dependence on the
interface thickness and for given $d_i$, either coupled or decoupled behavior can be
expected for all fields, as is indeed observed. \\
With the expression for the Josephson coupling force, it is simple to show that the
electromagnetic coupling force between two vortex segments can be neglected in our
considerations. It was shown by Clem \cite{clem75} that, for two superconductors
characterized by penetration depths $\lambda_{1,2} = \lambda$, with thickness
$d_{1,2} = d$ ($d \ll \lambda$), separated by an insulating layer of thickness $d_i$
(($d_i \ll \lambda$), the {\it maximum} force $F_{em}$ as function of the
displacement of the vortex segments with respect to each other is given by
\begin{equation} \label{eq:em}
F_{em} = \frac{\Phi_0^2}{8 \pi \mu_0} \left(\frac{d}{\lambda^2}\right)^2 .
\end{equation}
Here we neglected a weak field dependence which occurs in the regime of interacting
vortices, which would lower $F_{em}$ further. Writing Eq.~\ref{eq:em} as $F_{em} =
\epsilon_0 d^2 / (2 \lambda^2)$, we find that $F_{em} / F_J$ = $\gamma d^2 / ( 2
\lambda^2)$. With $d$ of order 50~nm and $\lambda$ of order 500~nm for both NbN and
\moge, $F_{em}$ is an order of magnitude lower than $F_J$ at the thickness $d_i$
where $F_{sh}$ starts to win from $F_J$. We should note, however, that this may not
be the case for superconductors with smaller values for $\lambda$.
\section{Conclusion}
In conclusion, we have used the technique of vortex core imaging by low-temperature
STM to study the restoring of initial disorder in a vortex lattice. The disorder was
engineered by depositing a weakly pinning superconducting thin film of \moge on a
strongly pinning layer of NbN. By varying the thickness of the weakly pinning layer,
and the interface conditions between both layers, we could reach several conclusions
with respect to the healing mechanisms. For clean interfaces we found that the
thickness where reordering of the vortex lattice is first observed can be linked to
the first bending mode of the vortex lines, without the need to renormalize the
elastic constants in the presence of such strong disorder. For our \moge~layers,
this thickness is around 75~nm. We then used layers with a thickness well below this
value in order to investigate the restoration of the vortex lattice when making the
interface more insulating. Here we find that the results can be understood in a
straightforward way by comparing the Josephson force working to align vortex
segments over the interface with the elastic restoring forces inside the weakly
pinning layer.

\section{Acknowledgement}
This work is part of the research program of the "Stichting voor Fundamenteel
Onderzoek der Materie (FOM)", which is financially supported by NWO. We are grateful
to A. E. Koshelev for making us aware of the results of ref.~\cite{goldschmidt05}.


\begin{thebibliography}{99}
\bibitem{hess89} H. F. Hess, R. B. Robinson, R. C. Dynes, J. M.
Valles and J. V. Waszczak, Phys. Rev. Lett. {\bf 62}, 214 (1989).
%
\bibitem{troyan99} A. M. Troyanovski, J. Aarts and P. H. Kes,
Nature {\bf 399}, 665 (1999).
%
\bibitem{renner98} Ch. Renner, B. Revaz, K. Kadowaki, I.
Maggio-Aprile and \O. Fischer, Phys. Rev. Lett. {\bf 80} 3606 (1998).
%
\bibitem{fischer06} \O. Fischer et al., Rev. Mod. Phys.
%
\bibitem{baarle03} G. J. C. v. Baarle, A. M. Troianovsky, T.
Nishizaki, P. H. Kes and J. Aarts, Appl. Phys. Lett. {\bf 82},
1081 (2003).
%
\bibitem{pruymboom87} A. Pruymboom, W.H.B. Hoondert, H.W. Zandbergen
and P.H. Kes, Jpn. Jn. Appl. Phys. \bf{26}\rm, Suppl. 26-3, 1529
(1987).
%
\bibitem{note-Rc} $r_c$ is equivalent to the 2D correlation length in the vortex lattice
for the overlapping vortex regime ($b$ $>$ 0.3; see E. H. Brandt, Phys. Rev. Lett.
{\bf 57}, 1347 (1986))
%
\bibitem{frankenthal83} R.P. Frankenthal, D.J. Siconolfi, W.R. Sinclair,
D.D. Bacon, J. Electrochem. Soc. \bf{130}\rm, 2056 (1983).
%
\bibitem{brandt76} E.H. Brandt, Phys. Status Solidi B{\bf 77}, 551 (1976).
%
\bibitem{goldschmidt05} Y. Y. Goldschmidt and S. Tyagi, Phys. Rev. B {\bf 71},
014503 (2005).
%
\bibitem{koshelev98} A.E. Koshelev, V.M. Vinokur, Phys. Rev. B
\bf{57}\rm, 8026 (1998).
%
\bibitem{note-model} In ref.~\cite{goldschmidt05} the case is considered of a kink in a line of
pancake vortices, and a correction is proposed to the expression for the Josephson
energy in order to take proper account of the fact that the line is not stiff and
the kink will be spread over several pancake segments. In our case the line segments
are stiff, and the correction would not be appropriate.
%
\bibitem{white94} W. R. White, A. Kapitulnik and M. R. Beasley, Phys. Rev. B {\bf
50}, 6303 (1994).
%
\bibitem{clem75} J. R. Clem, Phys. Rev. B {\bf 12}, 1742 (1975).
%
\end{thebibliography}
\end{document}